\documentclass[sigconf,pbalance]{acmart}
\AtBeginDocument{%
  }

\usepackage{lineno}
\usepackage{balance} 
\usepackage{caption}
\begin{document}
\setlength{\parskip}{0pt}
\title{Design and Evaluation of Whole-Page Experience Optimization for E-commerce Search}


\author{Pratik Lahiri}
\affiliation{%
  \institution{Amazon}
  \city{Seattle}
  \state{WA}
  \country{USA}
}
\email{lahirip@amazon.com}

\author{Bingqing Ge}
\affiliation{%
  \institution{Amazon}
  \city{Seattle}
  \state{WA}
  \country{USA}
}
\email{bqge@amazon.com}

\author{Zhou Qin}
\affiliation{%
  \institution{Amazon}
  \city{Seattle}
  \state{WA}
  \country{USA}
}
\email{qinzq@amazon.com}

\author{Aditya Jumde}
\affiliation{%
  \institution{Amazon}
  \city{Seattle}
  \state{WA}
  \country{USA}
}
\email{adijumde@amazon.com}

\author{Shuning Huo}
\affiliation{%
  \institution{Amazon}
  \city{Seattle}
  \state{WA}
  \country{USA}
}
\email{shuningh@amazon.com}

\author{Lucas Scottini}
\affiliation{%
  \institution{Roblox Corporation}
  \city{San Francisco}
  \state{CA}
  \country{USA}
}
\email{lcostascottini@gmail.com}
\authornote{Work done at Amazon}

\author{Yi Liu}
\affiliation{%
  \institution{Amazon}
  \city{Seattle}
  \state{WA}
  \country{USA}
}
\email{yiam@amazon.com}

\author{Mahmoud Mamlouk}
\affiliation{%
  \institution{Amazon}
  \city{Seattle}
  \state{WA}
  \country{USA}
}
\email{mmamlk@amazon.com}

\author{Wenyang Liu}
\affiliation{%
  \institution{Amazon}
  \city{Seattle}
  \state{WA}
  \country{USA}
}
\email{lwenyang@amazon.com}

\renewcommand{\shortauthors}{Pratik Lahiri et al.}
\renewcommand{\shortauthors}{Designing e-Commerce Search Whole Page Customer Experience Metrics}

\begin{abstract}
  E-commerce Search Results Pages (SRPs) are evolving from linear lists to complex, non-linear layouts, rendering traditional position-biased ranking models insufficient. Moreover, existing optimization frameworks typically maximize short-term signals (e.g., clicks, same-day revenue) because long-term satisfaction metrics (e.g., expected two-week revenue) involve delayed feedback and challenging long-horizon credit attribution. To bridge these gaps, we propose a novel Whole-Page Experience Optimization Framework. Unlike traditional list-wise rankers, our approach explicitly models the interplay between item relevance, 2D positional layout, and visual elements. We use a causal framework to develop metrics for measuring long-term user satisfaction based on quasi-experimental data. We validate our approach through industry-scale A/B testing, where the model demonstrated a 1.86\% improvement in brand relevance (our primary customer experience metric) while simultaneously achieving a statistically significant revenue uplift of +0.05\%. 
\end{abstract}





\maketitle

\section{Introduction}
E-commerce Search Results Pages (SRPs) have evolved from simple ranked lists into dynamic, visually complex layouts \cite{Lahiri2024, qin2024cooperativemultiagentdeepreinforcement}. As shown in Figure 1, for the query "XYZ water bottle," modern SRPs interleave organic search results (displayed in a grid) with themed widgets (e.g., "Trending Now"), each possessing distinct visual styles. This shift fundamentally alters user interaction: visual design drives non-sequential attention patterns that differ significantly from traditional top-to-bottom list scanning \cite{nielsen2006eyetracking}. Consequently, without changing the underlying item set, varying the placement of widgets creates distinct page layouts and thus distinct user experiences.

\begin{figure}[H]
    \centering
    \includegraphics[scale=0.5]{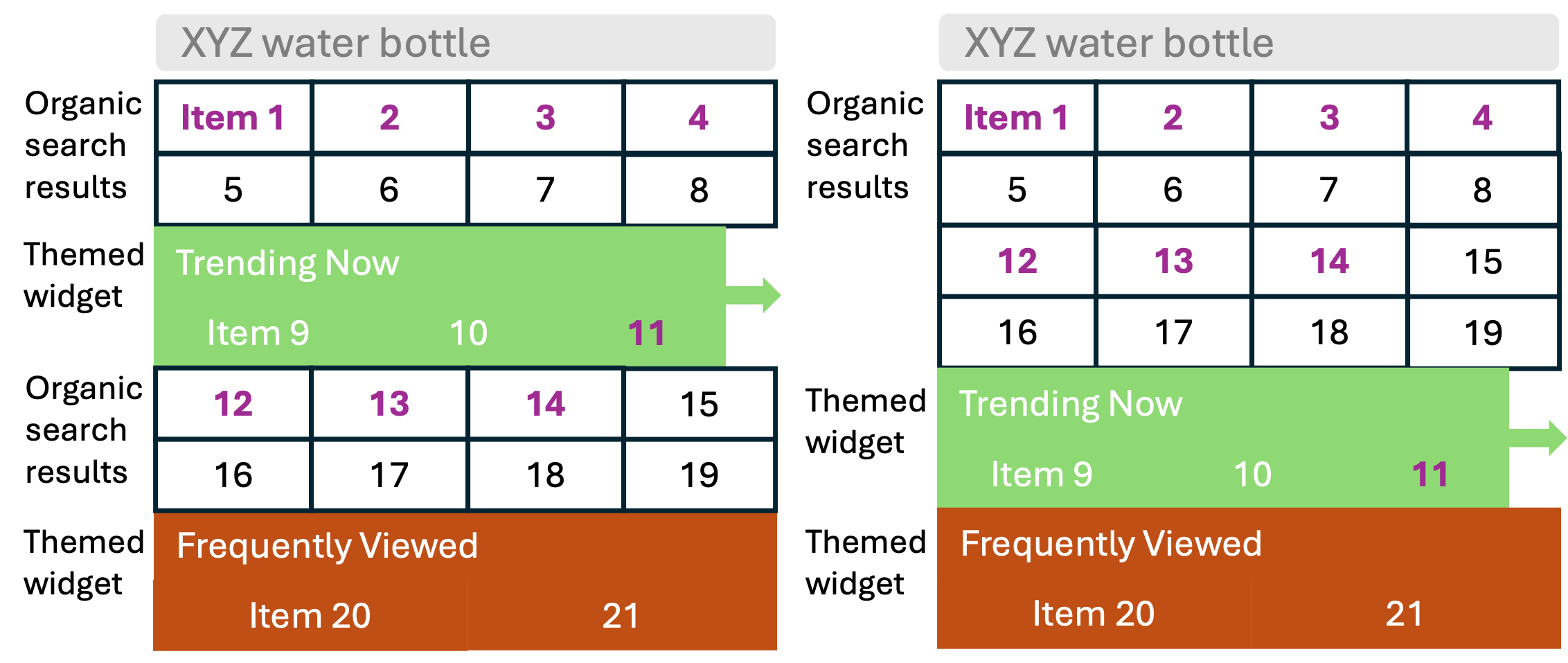}
    \label{fig:page_templates}
    \caption{Different search results page layouts. Brand XYZ items are shown in bold purple.}
\end{figure}
To manage the trade-off between exploring new layouts and exploiting known ones, bandit algorithms are widely used \cite{liu2021map, kawale2019netflix, 10.1145/3394486.3403374, mavridis2020beyond, Hill_2017, mao2019batchedmultiarmedbanditapproach}. Yet, applying slate optimization methods  \cite{10.1145/963770.963776, Hill_2017, ermis2020learningrankpositionbased, Qin2014ContextualCB} to this domain reveals four gaps: (1) \textbf{Multi-Objective Nature}: The system must balance engagement, revenue, and long-term user satisfaction; and (2) \textbf{Delayed Reward Signals}: success metrics, such as shopping frequency and long-term spend, are not observable on the same day; (3) \textbf{Complex Satisfaction Functions}: User satisfaction in 2D grids depends on the interplay of content and position, not just a simple discount factor; (4) \textbf{Heterogeneous Content}: Perception varies between organic search results and widgets.

We formulate SRP optimization as a contextual bandit problem, where the action is the presented SRP and contextual features are used to characterize content heterogeneity. To balance between multiple objectives, we propose a composite scalar reward defined as a weighted sum of page non-abandonment, short-term revenue, and a designed long-term impact based user satisfaction metric. Most prior work rewards SRP optimization using clicks \cite{10.1145/3292500.3330744, mavridis2020beyond, wsdm16} or short-term revenue \cite{www18, sigir18}. Some studies consider user satisfaction \cite{Chuklin_2016, 10.1145/1840784.1840801}, all measured via human annotation which suffers from scalability issues, annotator noise, and potential mismatch between annotation and real user preference \cite{10.1145/3204461, 10.1145/2766462.2767721, 10.1145/2766462.2767695, 10.5555/1631268}. We hypothesize that presenting high-quality SRPs generates positive long-term effects independent of immediate clicks or purchases, such as increased return visits and shopping frequency, ultimately leading to a higher long-term spend. We therefore use an estimate of expected long-term spend as a metric to quantify user satisfaction.
Our contributions are threefold:
\begin{enumerate}
    \item We propose a causal framework that uses quasi-experimental variation to link whole-page quality to long-term customer satisfaction, measured 12-week spend.
    \item We model position-dependent user sensitivity by partitioning the SRP into regions, and estimating region-specific user satisfaction, rather than assuming a continuous decay.
    \item We validate our method through online A/B testing with three treatments: no user satisfaction in the reward, click-through rate as a proxy, and our proposed satisfaction metric, which achieves the best performance.
\end{enumerate}
\section{Methodology}

\subsection{DV-WPX: A Causal Inference Framework}  
We propose DV-WPX (Downstream Value of Whole Page Experience), a causal framework that addresses \textbf{delayed reward signals gap} by linking observable whole-page quality to long-term customer satisfaction via quasi-experimental variation.
The model's identification strategy rests on the assumption that, conditional on historical customer features, variations in quality metrics across search events within a query (defined as keyword, search index alias, and ranking function combination), event date, and delivery locations are as good as experimental. This assumption holds exactly when variations stem from experimental assignments and approximately when variations arise from natural experiments like supply side shocks that are conditionally independent from unobserved customer characteristics driving downstream revenue (rev). The model approximates a setting where two equivalent shoppers (A and B) issue identical search queries on the same day from the same delivery locations, but experience different levels of search quality. This difference could emerge from experimental assignments or natural variations in search results. The model then estimates the difference in 12-week revenue between these shoppers after accounting for differences in short-term outcomes.
Quality metrics (except pricing and duplication) are analyzed across distinct page regions which can be adjusted (like a hyperparameter). This regional analysis strategy accounts for the varying impact of quality metrics based on their visibility and position in search results. The DV-WPX model's theoretical foundation rests on a structural equation that links downstream customer engagement to search quality and short-term behavior. For a search event e, the relationship is formalized as:
\[
rev_{t(e) + \delta_L} = W\big(rev_{t(e) + \delta_S}(Q_e),\ A_{t(e) + \delta_S}(Q_e),\ Q_e\big)
\]

where $rev_{t(e) + \delta_L}$ is cumulative long-term revenue (12 weeks post-search), 
$rev_{t(e) + \delta_S}$ is short-term revenue (2 weeks post-search), 
$A_{t(e) + \delta_S}$ denotes engagement metrics, 
and $Q_e$ is the vector of quality metrics. 
The welfare function $W(\cdot)$ maps short-term signals to long-term outcomes.

The causal effect of quality changes is captured through the total derivative.
\[
\frac{\partial rev_{t(e) + \delta_L}}{\partial Q_e} =
\frac{\partial W}{\partial rev_{t(e) + \delta_S}} \frac{\partial rev_{t(e) + \delta_S}}{\partial Q_e}
+ \frac{\partial W}{\partial A_{t(e) + \delta_S}} \frac{\partial A_{t(e) + \delta_S}}{\partial Q_e}
+ \frac{\partial W}{\partial Q_e}
\]

This decomposition reveals three distinct channels through which search quality affects downstream revenue. The first term captures how quality changes influence downstream revenue through their immediate impact on short-term sales. The second term represents the indirect effect through short-term engagement metrics. The final term measures the direct effect of quality on long-term outcomes, independent of its short-term impacts.

To estimate these effects empirically, the model employs a linear specification.
\[
Drev^{t}_{i,c,q,z} =
\sum_{s=1}^{S} \beta^{t}_{s} X^{t,s}_{i,c,q,z}
+ \sum_{j=1}^{J} \theta^{t}_{j} M^{j}_{i,c}
+ \sum_{k=1}^{K} \gamma^{t}_{k} H^{k}_{i,c}
+ \alpha^{t}_{q} + \zeta^{t}_{z} + \epsilon^{t}_{i,c,q,z}
\]
with corresponding surrogate equations:
\[
X_{i,c,q,z}^{t,s} = \sum_{j=1}^{J} \phi_{j}^{t,s} M_{i,c}^{j} + \sum_{k=1}^{K} \tau_{k}^{t,s} H_{i,c}^{k} + \alpha_{q}^{t,s} + \zeta_{z}^{t,s} + \epsilon_{i,c,q,z}^{t,s}
\]

Here, $i$ indexes search events, $c$ represents customers, $q$ denotes query characteristics, and $z$ indicates delivery ZIP codes. The $ \beta $ coefficients capture the causal effects of quality metrics (X), while $\theta$ and $\gamma$ represent the effects of short-term metrics (M) and historical controls (H), respectively. Fixed effects $\alpha_{q}$ and $\zeta_{z}$ account for query-specific and geographical variations.

The estimation employs Double Machine Learning (DML) \cite{chernozhukov2024doubledebiasedmachinelearningtreatment} to address potential confounding and ensure robust inference. The DML process consists of three carefully designed stages. First, the 0th stage implements an iterative de-averaging algorithm that removes fixed effects across queries and ZIP codes. This process requires 20 iterations to ensure convergence of residuals and downstream estimates. The data is then split into a 90 percent training fold and 10 percent testing fold to enable out-of-sample validation. The first stage of DML focuses on obtaining unbiased residuals of the de-averaged versions of both target and surrogate metrics. This is accomplished through linear ML models with two-fold cross-fitting, a technique that prevents overfitting by ensuring that the residuals for each observation are computed using models trained on different subsets of the data. The feature set includes comprehensive historical metrics and auxiliary controls. The second stage represents the core of the causal inference, implementing either OLS or LASSO regression of the residualized target metric on residualized surrogates. For LASSO estimation, the model employs sophisticated hyperparameter tuning through a 20-point grid search with 3-fold cross-validation. This ensures improved model selection while maintaining computational feasibility. The final DV-WPX metric is computed as:
\[
DVWPX_{i}^{t,E} = \sum_{s=1}^{S} \beta_{s}^{t,E} X_{i}^{t,s}
\]

This formulation aggregates the estimated causal effects ($\beta^t_s$) of individual quality metrics ($X_i$,$c$,$q$,$z$) into a single, interpretable measure of search quality's impact on long-term customer engagement.

\subsection{User Satisfaction Metric Derived from DV-WPX} 
The DV-WPX framework provides a general mechanism for translating observable whole-page quality signals into long-term user satisfaction. While agnostic to the specific quality signal, it enables the construction of satisfaction metrics whose components are weighted by their downstream impact rather than short-term interactions. Here, we instantiate DV-WPX using brand alignment on brand-sensitive queries as a concrete example to evaluate long-term satisfaction optimization in an online ranking system.

We design the Pixel and Region Weighted Whole-Page Brand Match Rate (PR-WP-BMR), which measures how well page content aligns with a branded query (e.g., ``Nike shoes''). The metric aggregates brand match rates across all items on the page, weighting each item by visual prominence (i.e., pixel coverage) and coarse page position (i.e., page region). 

To capture visual effects, we partition the SRP into three regions, regardless of whether items appear as standalone results or within widgets: Top (positions 1–8), Middle (positions 9–16), and Bottom (positions beyond 16). PR-WP-BMR is computed as a weighted sum of region-level, pixel-weighted brand match rates:
\[
\text{PR-WP-BMR}
= w_{\text{Top}} \cdot \overline{\text{BMR}}_{\text{Top}}
+ w_{\text{Mid}} \cdot \overline{\text{BMR}}_{\text{Mid}}
+ w_{\text{Bot}} \cdot \overline{\text{BMR}}_{\text{Bot}},
\]
where $w_{\text{Top}} + w_{\text{Mid}} + w_{\text{Bot}} = 1$ are weights for different regions.

A key design choice is how to set the region weights $\{w_r\}$. As a baseline, we derive weights from the empirical distribution of click-through rates (CTR) across page positions, reflecting short-term engagement patterns.
Alternatively, we derive region weights using DV-WPX, which estimates the marginal downstream value of an incremental brand match in each region. These effects are normalized to obtain relative weights that capture the long-term importance of brand alignment at different page locations. Empirically, DV-WPX assigns most weight to the Top and Middle regions, with negligible weight on the Bottom. Specifically, we use CTR-based weights $(0.60, 0.25, 0.15)$ and DV-WPX-based weights $(0.63, 0.37, 0)$ for the Top, Middle, and Bottom regions, respectively.

By combining pixel-based visual prominence with region-level weighting, PR-WP-BMR models satisfaction as a function of both layout and content, addressing \textbf{complex satisfaction functions} beyond a simple 1D position discount.

\subsection{Page Template Ranker}
To evaluate the DV-WPX-derived user satisfaction metric online, we integrate it into a production page template ranker.

\subsubsection{Problem Formulation}
In an industrial page template recommender, business constraints restrict how content can be displayed. There are $C$ possible ways to interleave themed widgets with search results, but not all arrangements are eligible due to business rules and UI constraints. We define a page template for each eligible way to rank content—two examples are shown in Figure 1. Let $P = \{P_1, \ldots, P_k\}$ denote the set of page templates, where each template $P_i$ admits a specific subset of eligible items $C_i \subseteq C$.

The action space is $A = \{a_1, \ldots, a_k\}$, where each action $a_i$ corresponds to selecting a page template and ordering its eligible items. Given an objective function $F : A \rightarrow \mathbb{R}$ that predicts the reward of an action, the ranker selects the action that maximizes $F$. In practice, this corresponds to choosing, for each search request, a page template from the pool of eligible templates $P$.

Based on business needs, our problem is to learn a decomposition of $F$ that optimizes for multiple objectives, including long-term delayed rewards.

\subsubsection{Input Features}
The ranker uses a rich set of input signals, referred to as the 3Cs: Context, Customer, and Content. Context features include, but are not limited to, marketplace, device type, inferred query specificity, and product categories at multiple granularities. Customer features capture status-related signals such as membership. Content features describe the relevance and value of each rankable, including aggregated measures of relevance, brand alignment, product-type alignment, and other value-related signals. These features are computed in a content-type-aware manner (e.g., separate aggregation and calibration for search result vs. widgets), allowing the model to capture different user responses to \textbf{heterogeneous content}.

\subsubsection{Modeling}
The ranker performs multi-objective optimization over business outcomes (i.e., revenue), engagement, and whole-page satisfaction. Engagement is measured by binary page non-abandonment, while whole-page satisfaction is captured through the metric defined in the previous section (i.e., PR-WP-BMR).


We train a separate predictive model for each objective. Continuous objectives use Bayesian linear regression \cite{blir}, while binary objectives use Bayesian probit regression \cite{blip}. Training uses historical impression logs, with inputs consisting of the 3Cs features and the displayed template, and targets corresponding to each objective. Linear regression models minimize root mean squared error, while probit models maximize area under the ROC curve. Models are refreshed daily using incremental training, sampling 50\% of data from the most recent day to adapt to evolving traffic and content.


\subsubsection{Inference and Decision Process}
At inference time, the ranker evaluates all candidate templates and applies Thompson sampling to generate predictions for each objective. For each template and context, samples are drawn from the posterior distributions of the corresponding objective models. The sampled predictions are combined into a single reward score using a weighted linear combination of objectives. The weights are determined offline based on the historical statistics (e.g., mean and variance) of each objective to normalize their scales and ensure comparable contribution to the final score. The template with the highest aggregated score is selected for display. This scalar reward provides a practical mechanism and addresses the \textbf{multi-objective nature} of template selection.

\section{Results}
We evaluate the impact of incorporating user satisfaction into the page template ranker through offline and online experiments. We consider three settings: a Control without any satisfaction signal, Treatment~1 (T1) using a CTR-based PR-WP-BMR proxy, and Treatment~2 (T2) using a DV-WPX-based PR-WP-BMR to optimize long-term satisfaction. In both treatments, models also incorporate content-aware features capturing query–widget alignment in relevance, brand, and product type.

We present offline results to evaluate content-aware features improve predictions on objectives shared across models. Since PR-WP-BMR is absent from the Control, offline comparisons focus on common revenue and engagement metrics. Because T1 and T2 share the same models for these objectives and differ in the satisfaction metric, their offline results are identical and reported jointly.

\subsection{Offline Data and Results}
In both the Control and Treatment settings, separate models are trained to predict revenue and non-abandonment. Offline training and evaluation use one day of North America traffic, comprising approximately 1.5 million Desktop impressions and 5.6 million Mobile impressions, with a 99\% / 1\% train–test split.

Model performance on the held-out test set is reported in Table~\ref{tab:offline_results}. We use root mean squared error (RMSE; lower is better) for continuous outcomes and area under the ROC curve (AUC; higher is better) for binary outcomes; positive percentage changes indicate improvement. Non-abandonment is included as an optimization objective only for Desktop in our MOO formulation, so we report it only for Desktop.

The results show that adding content-aware features improves revenue prediction on both Mobile and Desktop, and slightly improves non-abandonment prediction on Desktop. 

\begin{table}[h]
\small
\centering
\caption{Offline performance comparison between baseline and treatment models.}
\label{tab:offline_results}
\begin{tabular}{llrrrr}
\hline
Dev. & Metric & Eval. & Base. & Treat. & $\Delta$\% \\
\hline
Mob  & Revenue & RMSE & 47.70 & 44.90 & +6 \\
Desk & Non-Aband. & AUC & 0.39 & 0.39 & +1 \\
Desk & Revenue & RMSE & 8.87 & 8.32 & +6 \\
\hline
\end{tabular}
\end{table}



\subsection{Online Test Results}

We also evaluate the changes in a one-month worldwide A/B test on both Mobile and Desktop. Table~\ref{table:metrics} reports the rolled-up online results, comparing each treatment variant against the Control.

Incorporating a user satisfaction signal into the page template ranker improves customer experience and business outcomes regardless of how region weights are defined. Both T1 (CTR-based PR-WP-BMR) and T2 (DV-WPX-based PR-WP-BMR) deliver positive short-term revenue gains of 0.04\% and 0.05\%, respectively.

Notably, using DV-WPX-derived weights (T2) yields additional benefits beyond short-term performance. While T1 shows a slight decline in long-term revenue, T2 achieves a positive lift, indicating that DV-WPX better captures signals aligned with long-term customer satisfaction and value.

Although DV-WPX weights are not derived from click behavior, both T1 and T2 improve engagement, as reflected by identical gains in search CTR. This suggests that optimizing for whole-page satisfaction does not trade off short-term engagement, and that DV-WPX-based weighting can improve long-term outcomes without sacrificing immediate user interaction.

\begin{table}[h]
\centering
\small
\caption{Online performance comparison. All reported lifts are statistically significant.}
\label{table:metrics}
\begin{tabular}{lcc}
\hline
Metric & Relative Lift (T1--C) & Relative Lift (T2--C) \\
\hline
Revenue            & 0.04\%   & 0.05\%   \\
Long-term Revenue  & -0.001\% & 0.005\%  \\
Search CTR         & 0.02\%   & 0.02\%   \\
\hline
\end{tabular}
\end{table}


\section{Discussion}

This work studies how to incorporate long-term user satisfaction into whole-page optimization for e-commerce search. We introduce DV-WPX, a causal framework that maps whole-page quality to downstream value, and instantiate it with a brand satisfaction metric (PR-WP-BMR). We integrate this metric into a production page template ranker and evaluate it offline and online. Results show that satisfaction-aware optimization improves performance, with DV-WPX-based weighting providing additional long-term benefits beyond short-term engagement proxies. 

While the results are encouraging, several directions remain for future work. First, although we focus on brand relevance as a concrete instantiation, the DV-WPX framework can be extended to other whole-page quality dimensions such as visual diversity or cross-component coherence. Second, the current DV-WPX model relies on a 12-week observation window, which may be long for rapidly evolving retail settings; exploring shorter or adaptive horizons is an important next step. Finally, while we use fixed positional regions in this work, future extensions could consider dynamic region definitions that adapt to device form factors and user interaction patterns. Despite these opportunities, our approach demonstrates a practical and scalable solution that has been successfully deployed in a real-world e-commerce system.

\section{Ethical Considerations}
This paper presents work whose goal is to advance the field of Machine Learning. There are many potential societal consequences of our work, none which we feel must be specifically highlighted here.

\bibliographystyle{ACM-Reference-Format}
\bibliography{paper-reference}

\end{document}